# Atomically Thin Metal-Dielectric Heterostructures by Atomic Layer Deposition

Pallabi Paul[1,2,6] Paul Schmitt[1,2,6], Vilborg Vala Sigurjónsdóttir[1,2], Kevin Hanemann[2], Nadja Felde[2], Sven Schröder[2], Felix Otto[3], Marco Gruenewald[3], Torsten Fritz[3], Vladimir Roddatis[4,5], Andreas Tünnermann[1,2], Adriana Szeghalmi[1,2]*

[1] Institute of Applied Physics and Abbe Center of Photonics, Friedrich Schiller University Jena, Albert-Einstein-Straße 15, D-07745 Jena, Germany

[2] Fraunhofer Institute for Applied Optics and Precision Engineering IOF, Center of Excellence in Photonics, Albert-Einstein-Straße 7, D-07745 Jena, Germany

[3] Institute of Solid-State Physics, Friedrich Schiller University Jena, Helmholtzweg 5, 07743 Jena, Germany

[4] Institute of Materials Physics, University of Göttingen, Germany

[5] Current address: German Research Centre for Geosciences GFZ, Helmholtz Centre Potsdam, Germany

[6] These authors contributed equally to this work.

**ABSTRACT:** Heterostructures increasingly attracted attention over the past several years to enable various optoelectronic and photonic applications. In this work, atomically thin interfaces of Ir/$Al_2O_3$ heterostructures compatible with micro-optoelectronic technologies are reported. Their structural and optical properties were determined by spectroscopic and microscopic techniques (XRR, XPS, HRTEM, spectroscopic ellipsometry, and UV/VIS/NIR spectrophotometry). The XRR and HRTEM analyses reveal a layer-by-layer growth mechanism of Ir in atomic scale heterostructures, which is different from the typical island-type growth of metals on dielectrics. Alongside, XPS investigations imply the formation of Ir-O-Al bonding at the interfaces for lower Ir concentrations, in contrast to the nanoparticle core-shell structure formation. Precisely tuning the ratio of the constituents ensures the control of the dispersion profile along with a transition from effective dielectric to metallic heterostructures. The Ir coating thickness was varied ranging from a few Å to films of about 7 nm in the heterostructures. The transition has been observed in the structures containing individual Ir coating thicknesses of about 2-4 nm. Following this, show epsilon-near-zero metamaterials with tunable dielectric constants by precisely varying the composition of such heterostructures. Overall, a comprehensive study on structural and optical properties of the metal-dielectric interface of Ir/$Al_2O_3$ heterostructures was addressed indicating an extension of the material portfolio available for optical system design.

## INTRODUCTION

Due to the upsurge of exploring novel functionalities and continuously shrinking device sizes for 3D photonic integration, the investigation of various heterostructures, nanocomposites, and multilayers becomes increasingly demanding. Ultrathin metallic layers have been explored for applications in nanoelectronics, plasmonics, and photonics[1–3]. Metal-dielectric multilayer structures have revealed remarkable applications in enhancing optical nonlinearity[4], photocatalysis[5], and ultrafast transport of plasmonic electrons[6], to name only a few.

Such heterostructures were previously grown by evaporation[7] or sputtering[8] techniques exhibiting various optical applications [9], for instance, in surface enhanced Raman spectroscopy[10] and solar selective absorbers[11]. Here, we demonstrate the potential of atomic layer deposition (ALD) technology for realizing atomically thin metal-dielectric heterostructures. ALD is a powerful coating technology for developing novel nanostructured optical components due to the conformal growth of thin films on complex 3D architectures or high aspect ratio structures. The film growth is based on sequential self-limited surface reactions of a metal-organic precursor and an oxidizing or reducing agent[12]. This technique is approved for obtaining highly precise nanostructures of optically relevant materials which are otherwise difficult to implement[13–15].

Noble metal Ir exhibits a plethora of fascinating properties and applications, for instance, designing X-ray mirrors due to its high density[16–18], metal wire grid polarizers in the UV[13], microelectronics[19], and protection layers[20] are noteworthy. On the other hand, $Al_2O_3$, a very technologically relevant oxide represents wide range of functionalities ranging from passivation layer in photovoltaics[21], barrier coating[22], adhesion layer[23] to memory applications[24]. In this work, we investigate the structural, optical, and interfacial properties of heterostructures formed between Ir and $Al_2O_3$ grown by ALD. The heterostructures undergo an effective insulator-to-metal transition depending on the inclusion of Ir in the layer stacks. Further, the tunability of the optical dispersion profile can bring out novel functionality, such as, epsilon-near-zero (ENZ) behaviour. This might enable potential applications of multilayer films for enhancing nonlinear optical response. Altogether this work provides a basis to realize and explore the technological potential of novel atomically thin metal-dielectric heterostructures.

**EXPERIMENTAL DETAILS**

**Fabrication.** The $Ir/Al_2O_3$ heterostructures were fabricated on Si wafers and fused silica substrates with a SunALE R-200 (Picosun Oy, Masala, Finland) using $Ir(acac)_3$ and molecular oxygen ($O_2$) as precursors for Ir, and trimethyl aluminum (TMA) and water ($H_2O$) as precursors for $Al_2O_3$, respectively. The deposition temperature was kept at 380 °C. The corresponding ALD process parameters for the deposition of Ir and $Al_2O_3$ are as mentioned below
- 6 s $Ir(acac)_3$ pulse/60 s $N_2$ purge/2 s $O_2$ pulse/6 s $N_2$ purge
- 0.1 s TMA pulse/4 s $N_2$ purge/0.2 s $H_2O$ pulse/ 4 s $N_2$ purge

resulting in a growth per cycle (GPC) of 0.6 Å/cycle and 0.9 Å/cycle, respectively [25]. Upon developing single-layer Ir and $Al_2O_3$ ALD films, two sets of heterostructures were prepared. In these heterostructures, $Al_2O_3$ ALD cycles (i.e., the spacer layer thickness) were kept at 35 and 15 cycles, whereas the number of Ir ALD cycles has been varied as 4, 8, 16, 32, 64, and 128 cycles. These stacks were repeated to achieve at least a minimum of about 120 nm thickness in order to facilitate subsequent characterizations. The notation for identifying the samples is '$[Ir:Al_2O_3]*N$', where Ir and $Al_2O_3$ refer to the number of ALD cycles of each material and N is the number of repetitions. For example, a composition '[4:15]*102' consists of 4 Ir cycles followed by 15 $Al_2O_3$ cycles, along with 102 times repetition of this sequence.

**Characterizations.**
X-ray reflectivity (XRR) investigations were performed using an X-ray diffractometer D8 (Bruker-AXS D8 advanced system (Bruker AXS, Karlsruhe, Germany) in Bragg-Brentano geometry with a Cu Kα X-ray source (radiation wavelength 0.154 nm). For grazing incidence X-ray diffraction (GIXRD), a universal X-ray diffractometer URD65 (Freiberger Präzisionsmechanik, Freiberg, Germany) with Cu Kα radiation at a fixed angle of incidence of 0.3°, a grazing incidence collimator, and a detector with Soller collimation were used. The individual layer thickness within the heterostructures and the presence of an amorphous nature of the heterostructures were determined by XRR and GIXRD techniques, respectively.
The film composition was analyzed using X-ray photo-electron spectroscopy (XPS). Therefore, an XR 50 M X-ray source with FOCUS 500 monochromator (SPECS Surface Nano Analysis GmbH, Berlin, Germany) and Al Kα radiation (1486.7 eV) was used in normal emission at room temperature. The photoelectrons were acquired with a PHOIBOS 150 hemispherical analyzer equipped with a delay line detector (3D DLD4040-150).
Scanning electron microscope (SEM) images were obtained with a field emission SEM Hitachi S-4800 (Hitachi, Tokyo, Japan). Scanning tunneling microscopy (STM) was performed at room temperature using an Easyscan 2 STM (Nanosurf) operated with a Pt/Ir tip. For the white light interferometry measurements, a NewView 7300 system (Zygo, Middlefield, CT, USA) with 50x magnification was used to evaluate the surface roughness of the layer stacks.
High-resolution transmission electron microscopy (HR-TEM) (Thermo Fisher Scientific, formerly FEI, Eindhoven, Netherlands) investigations were performed using an image aberration-corrected Titan 80-300 environmental TEM operated at 300 kV and equipped with a Gatan imaging filter Quantum 965ER (Pleasanton, USA). The corresponding samples were prepared by mechanical polishing followed by $Ar^+$ milling using a Gatan PIPS 695 setup.
The film thicknesses, GPC, and dispersion profiles of the $Ir/Al_2O_3$ heterostructures were determined by means of spectroscopic ellipsometry (SE) using an SE850 DUV spectroscopic ellipsometer (Sentech Instruments GmbH, Berlin, Germany). SE measurements were performed in the wavelength range of 200-980 nm. A Drude model combined with 2-7 Lorentz oscillators was successfully incorporated to estimate the optical constants of such nanostructures. The transmission (T) and reflection (R) spectra were measured on fused silica substrates using a Lambda 900 spectrophotometer from PerkinElmer Inc. (Waltham, MA, USA) for 200-2000 nm wavelength. The optical losses (OL) were calculated as follows, OL (%) = 100-(T+R). These measurements were also used to verify the optical constants determined by spectroscopic ellipsometry.

**RESULTS AND DISCUSSION**

Several articles reported on the island-type nucleation behavior of noble metals instead of continuous layers [26–30]. The growth of Ir nanoparticles and ultrathin metallic layers by ALD was extensively studied and the Volmer Weber growth mode was observed during the nucleation regime [25,31,32]. Eventually, the Ir islands expand and approach closer to form a percolated network, subsequently leading towards the formation of closed metallic Ir layers. To further understand the evolution of surface coverage of only a few Ir ALD cycles, we performed STM measurements on a gold-coated mica substrate treated with 4 ALD cycles of Ir. Figure 1 demonstrates that the surface coverage after 4 ALD cycles of Ir is about 97%. After the very first ALD cycle, Ir covers almost the entire substrate, behaving as an initial wetting layer. Upon further increasing the number of ALD cycles up to 4, we observed a discrete accumulation of Ir on preferred substrate regions forming an island-like growth.

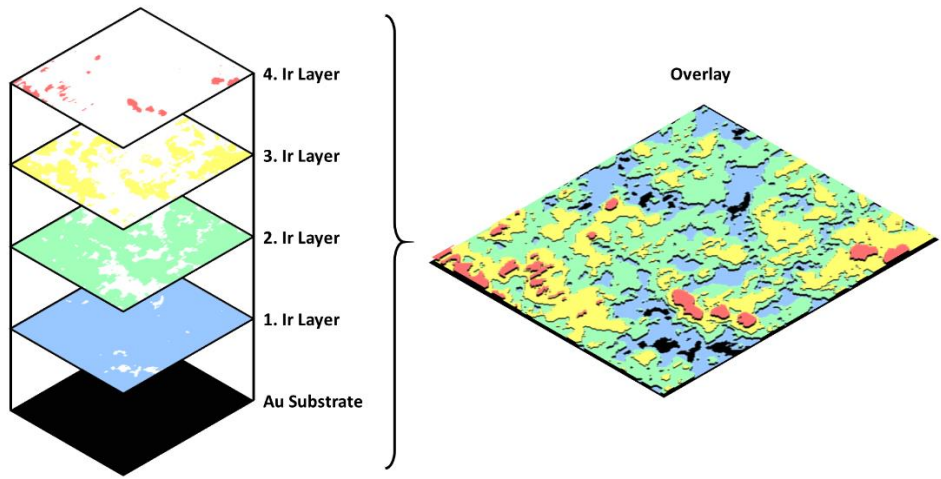

**Figure 1.** STM analysis (Voltage = 2.5 V and Current = 1.0 nA, measurements performed in constant current mode) of 4 ALD cycles of Ir deposited on a gold-coated mica substrate. This scan area is about 0.13 x 0.13 µm².

In order to elucidate the growth mechanism of ultrathin Ir layers in heterostructures, XRR measurements were conducted. For the Ir/$Al_2O_3$ heterostructures with a low number of Ir ALD cycles, Ir nanoparticles were presumed to be embedded in the $Al_2O_3$ matrix, based on the Volmer-Weber growth mechanism. On the contrary, our XRR analysis reveals the presence of a Bragg peak at about 2.8° even for 2 cycles of Ir, as illustrated in Figure 2 (a), indicating a typical layered heterostructure. This interpretation is even more evident from the HRTEM measurement on a [4:35]*50 heterostructure demonstrating a periodic bilayer structure of Ir and $Al_2O_3$. In Figure 2(b), the bright layers denote $Al_2O_3$ and the dark layers display Ir.

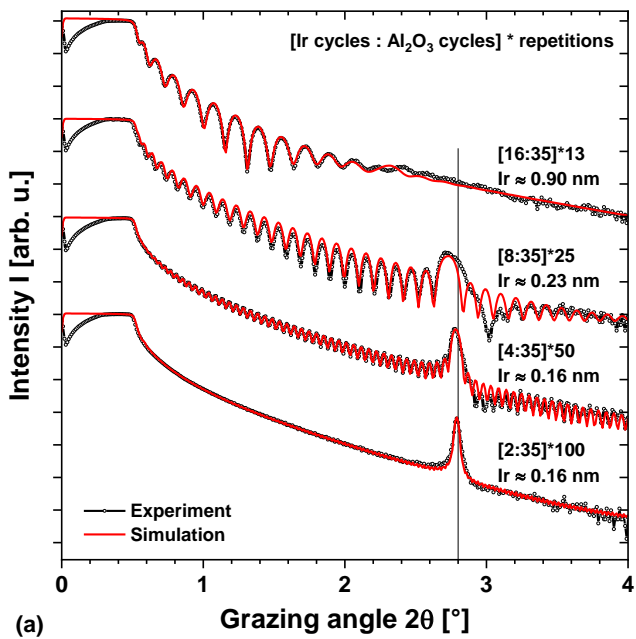
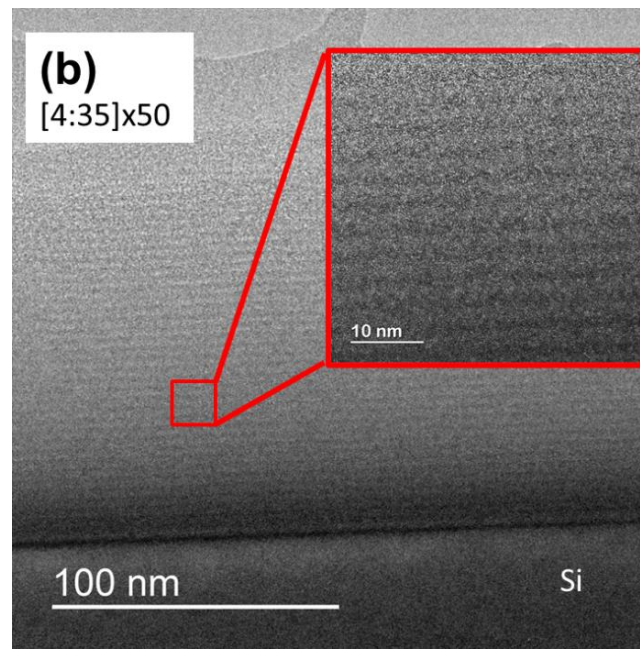

**Figure 2.** (a) XRR measurements with corresponding simulations of selected heterostructures, (b) HRTEM micrograph of a [4:35]*50 heterostructure sample.

In addition, grazing incidence XRD (GIXRD) measurements were performed on a selected set of heterostructures as illustrated in Figure 3. These heterostructures show no presence of an XRD peak indicating an amorphous growth of the Ir/Al$_2$O$_3$ nanostructures. Only for an ultrathin Al$_2$O$_3$ (5 ALD cycles ≈ 0.5 nm) individual layer, XRD peaks corresponding to Ir are visible. These Ir layers with 32 cycles (≈ 2 nm) are polycrystalline. All heterostructures with 35 cycles of Al$_2$O$_3$ barrier layers are amorphous in the GIXRD measurements. This spacer film thickness (≈ 3.1 nm) is sufficient to inhibit the crystallization of Ir.

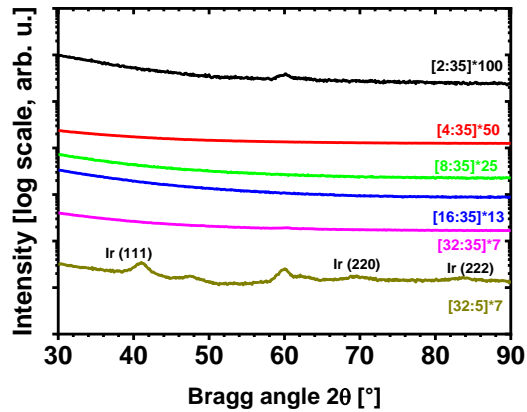

**Figure 3.** GIXRD measurements of selected Ir/Al$_2$O$_3$ heterostructures.

From the XRR and HRTEM investigations, an intermixing of Ir and Al$_2$O$_3$ at the interfaces has been concluded. To ensure a better understanding of the bonding environment and structural integrity for such ultrathin heterostructures, two samples were exclusively prepared to make them suitable for XPS measurements. Here, it was essential to obtain the extent of the metallicity of these nanostructures with atomically thin Ir contributions. The design of these structures was as follows,

- [Ir 2 cycles : Al$_2$O$_3$ 35 cycles]* 3 + Ir 2 cycles + Al$_2$O$_3$ 20 cycles
- [Ir 16 cycles : Al$_2$O$_3$ 35 cycles] * 3 + Ir 16 cycles + Al$_2$O$_3$ 20 cycles
- Single layer Ir – 200 cycles, as reference sample

In Figure 4(a), the heterostructures are denoted with '4' times repetition to keep the notation short and consistent as before. It was important to prepare the layer stacks with only a few bilayers and an ultrathin Al$_2$O$_3$ (20 ALD cycles ≈ 1.8 nm) layer on top to reduce charging effects (spectral shifts) in XPS. The top Al$_2$O$_3$ layer was made with 20 cycles instead of 35 cycles (≈ 3.1 nm) as the latter was too thick for the photoelectrons to escape due to their limited escape depth. Additionally, it was necessary to overcoat the top Ir layer with an ultrathin Al$_2$O$_3$ layer to incorporate a sandwiched Ir layer between the Al$_2$O$_3$ layers. A single layer Ir film (200 cycles ≈ 12 nm) has been investigated as a reference. In Figure 4(a), a bare Ir film depicted by the black curve displays the contribution from Ir 4f$_{7/2}$ and Ir 4f$_{5/2}$ states with an excellent agreement to the literature values [33–36]. The red curve illustrates a heterostructure with 16 cycles of Ir (≈ 0.9 nm), where the XPS profile corresponds very well with the pure Ir 4f states (indicated by the black curve). Mainly a metallic Ir bonding state is prevalent; however, a significant contribution towards higher binding energies can be attributed to the subtle presence of an Ir-O-Al bonding environment. A quantitative analysis shows an approximately 68% metallic and 32% oxidic chemical composition. This chemical composition can be interpreted as about 16% of oxidation along the Ir- Al$_2$O$_3$ interface on each side. For instance, in the case of the [16:35]*4 composition, about 1-2 monolayers of Ir are assumed to be oxidized along the interfaces with Al$_2$O$_3$ on both sides. A schematic diagram is in Figure 4(b) illustrates the formation of Ir-O bonding for very thin Ir layers in the heterostructure along the interfaces with Al$_2$O$_3$ on both sides. With further decreasing the Ir contribution (2 ALD cycles ≈ 0.1 nm), indicated by the blue curve, we obtain a significant shift towards higher binding energies indicating the presence of an Al-O-Ir-O-Al bonding state. The bond length of Ir-Ir is about 2.7 Å; therefore, compositions having 2, 4, and 8 cycles of Ir mainly consist of O-Ir-O bonded layers, whereas with 16 cycles of Ir (0.9 nm), 2-3 metallic stacks are sandwiched between Ir-O bonded monolayers. The bond-dissociation energies of Ir-Ir, Ir-O, Al-O are 361± 68 kJ/mol, 414±42 kJ/mol, and 501 ± 10 kJ/mol, respectively [37], indicating the possibility of Ir-O bond formation at the Ir/Al$_2$O$_3$ interfaces.

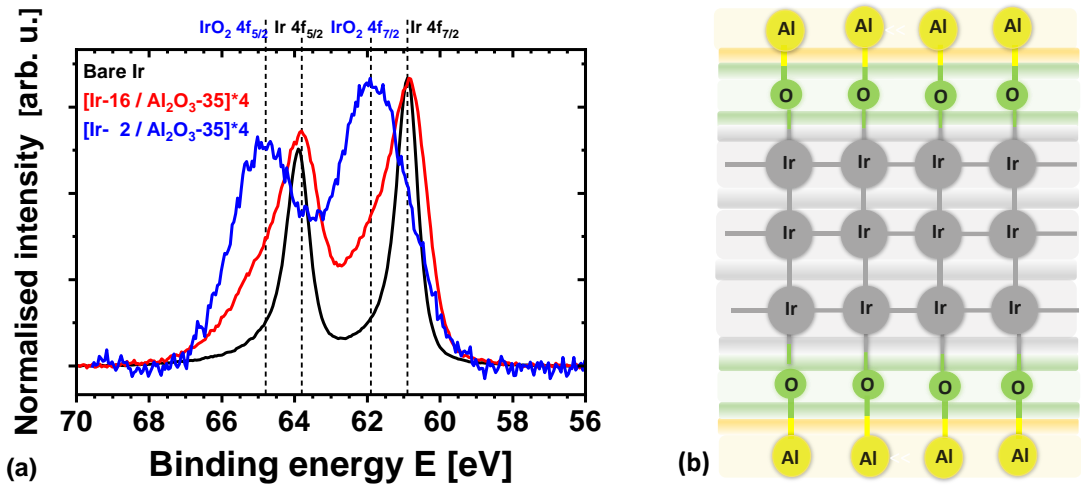

**Figure 4.** (a) XPS measurements of specially prepared Ir/Al$_2$O$_3$ heterostructures compared with a single-layer Ir film. (b) Schematic representation of atomically thin Ir layers in the heterostructure forming Ir-O bonding along the interfaces with Al$_2$O$_3$.

We further investigated the optical properties of these metal-dielectric interfaces by means of UV/VIS/NIR spectrophotometry and SE. Here, we studied the evolution in optical properties by systematically increasing the Ir content, along with two different Al$_2$O$_3$ spacer layer thicknesses. Table 1 contains detailed information about the calculated and determined total stack thicknesses employing SE, XRR, and SEM techniques along with the derived root mean square (r.m.s.) surface roughnesses of the compositions determined by the WLI measurements. The error in thickness determination by SE was about ±0.5 nm. For the samples with higher Ir contributions, meaning, 32, 64, and 128 ALD cycles, XRR measurements show a bad signal-to-noise ratio making them unsuitable for the fitting procedure.

**Table 1.** Determination of total stack thicknesses and r.m.s. surface roughnesses of ALD-grown Ir/Al$_2$O$_3$ heterostructures.

| Compositions [Ir:Al$_2$O$_3$]*N | Total stack thickness [nm] | | | | Roughness (r.m.s.) [nm] |
| --- | --- | --- | --- | --- | --- |
| | *Calculated* | *SE* | *XRR* | *SEM* | *WLI* |
| **[4:35]*102** | 343.7 | 335.9 | 326.4 | 312.3 | 1.1 |
| **[8:35]*102** | 366.2 | 330.3 | 324.2 | 320.6 | 0.4 |
| **[16:35]*102** | 411.1 | 351.9 | 345.8 | 395.2 | 0.5 |
| **[32:35]*51** | 250.4 | 226.5 | - | 282.5 | 1.1 |
| **[64:35]*25** | 166.8 | 152.8 | - | 202 | 0.4 |
| **[128:35]*13** | 132.5 | 129.4 | - | 156.7 | 1.1 |
| **[4:15]*102** | 160.1 | 140.4 | 136.4 | 137.0 | 0.6 |
| **[8:15]*102** | 182.6 | 139.6 | 147.2 | 146.5 | 1.0 |
| **[16:15]*102** | 227.5 | 153.5 | 213.5 | 224.7 | 1.3 |
| **[32:15]*51** | 158.6 | 128.6 | - | 194.5 | 0.7 |
| **[64:15]*25** | 121.8 | 108.1 | - | 159.5 | 0.7 |
| **[128:15]*13** | 109.1 | 116.1 | - | 137.5 | 1.1 |

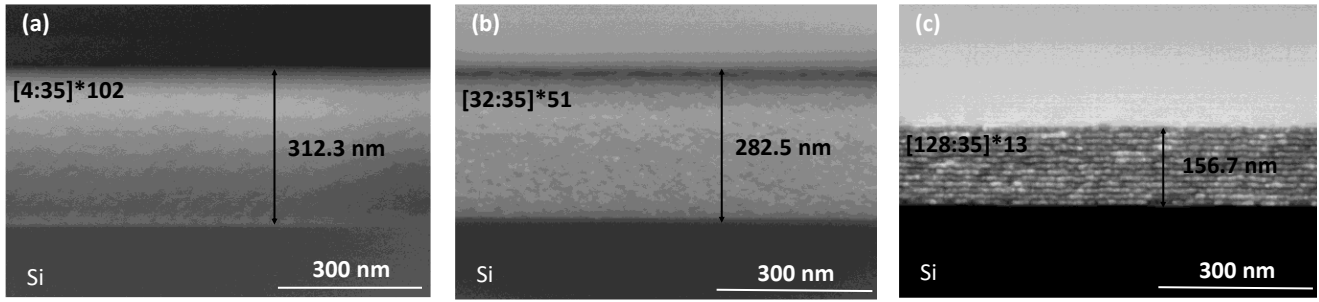

**Figure 5.** SEM micrographs of selected composites; (a) [4:35]*102, (b) [32:35]*51, (c) [128:35]*13. For very thin Ir coatings (i.e., up to 32 cycles ≈ 2 nm), SEM could not resolve the heterostructures. In the case of 128 Ir cycles (≈ 7.6 nm), a layered heterostructure is visible.

The measured T and R spectra of corresponding heterostructures are demonstrated in Figure 6 (a, b, d, e). The optical losses shown in Figure 6 (c, f) were calculated from the measured T and R data set. Figure 6 (a-c) illustrate the optical spectra of the heterostructures having about 3.1 nm $Al_2O_3$ spacer thickness (i.e., 35 ALD cycles). Initially, the transmission spectra are quite similar up to 16 cycles of Ir inclusion with a gradual decrease especially in the UV region. The transmission rapidly decreases for 32 cycles and higher, indicating a rise in the effective metallic character. With 128 Ir ALD cycles (≈ 7.6 nm) in the heterostructure, the stack becomes non-transparent owing to the formation of closed metallic layers creating a nanolaminate structure, as shown in Figure 5(c). Reflection spectra indicate the metallic character for the thickest Ir contribution.

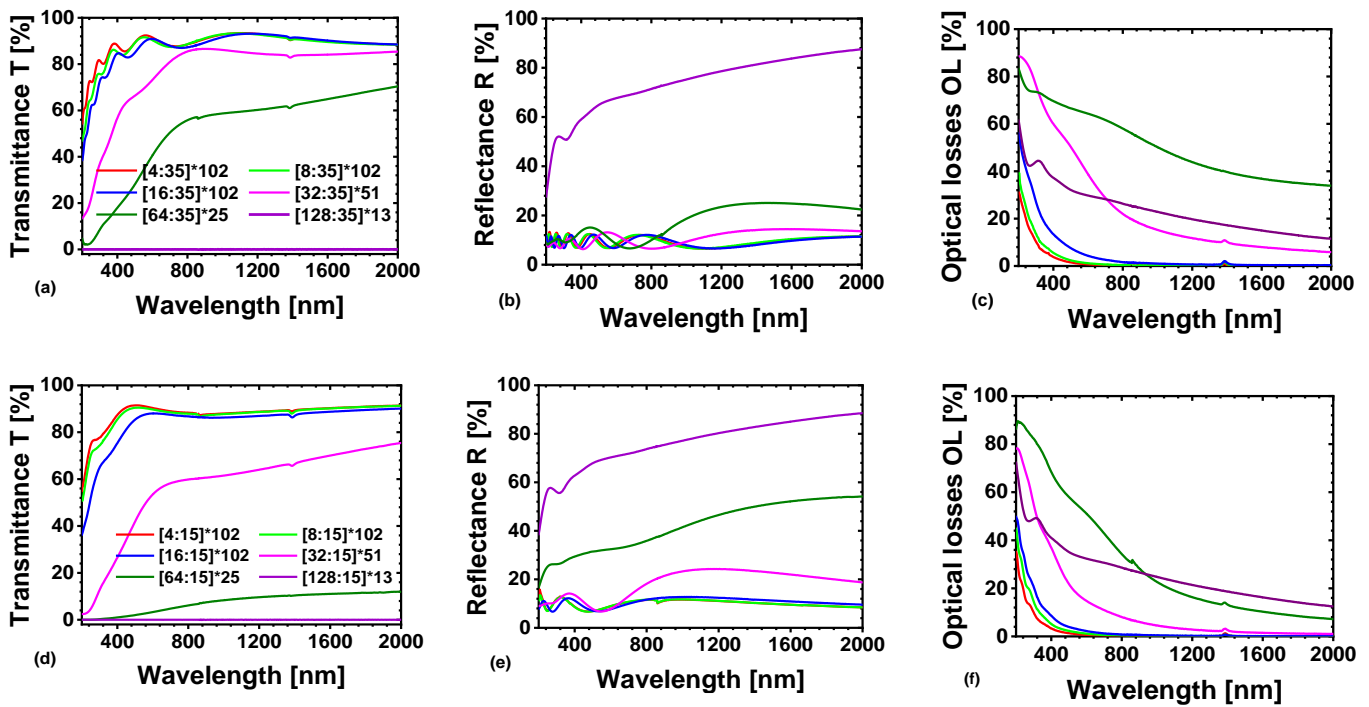

**Figure 6.** Measured transmittance T, reflectance R, and calculated optical losses OL spectra of Ir/$Al_2O_3$ heterostructures with (a, b, c) 35 cycles and (d, e, f) 15 cycles of $Al_2O_3$ spacer layers.

In Figure 6 (d-f), we depict the optical spectra for the samples with 1.4 nm $Al_2O_3$ spacer layer thickness (15 ALD cycles). The decrease in transmission was more rapid than for the systems having thicker spacer layers. In addition, the reflection spectra start indicating a metallic nature already at 64 Ir cycles. Metallicity emerged at a lower Ir content when the spacer thickness was smaller. However, for ultrathin Ir contributions (16 cycles or less), even a 1.4 nm spacer layer was a sufficient barrier to attain a partially metallic character of the stacks, which has been more evident from the optical constants determined by SE.

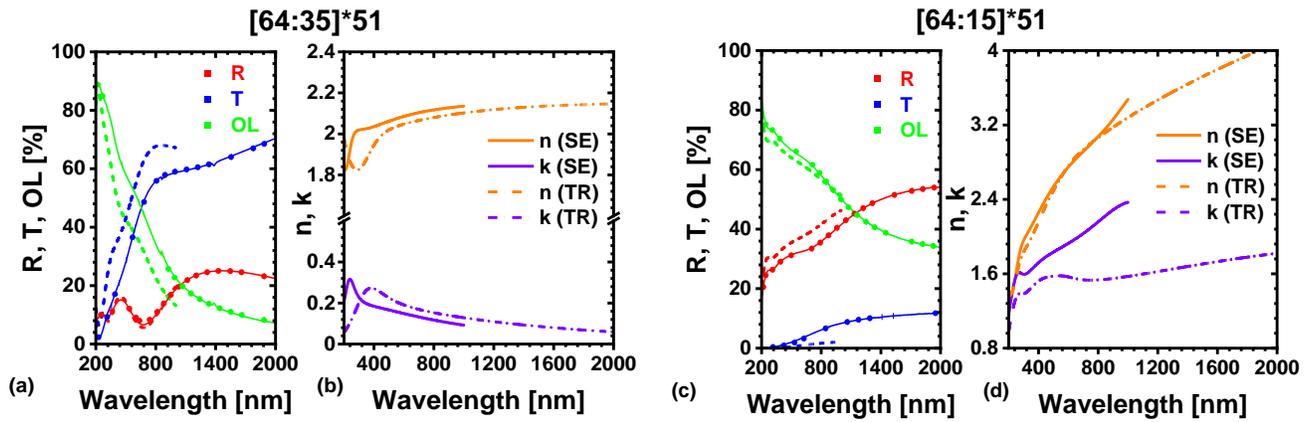

**Figure 7.** Model results for two composites. Three data sets for the reflectance R (red), transmittance T (blue) and corresponding optical losses OL (green) are compared (a, c); the measured T/R/OL spectra (solid lines), fits of the T/R/OL measurements (dots), and the T/R/OL spectra calculated from the oscillator model (dashed lines) used to fit the SE measurements. The refractive index (orange) and extinction coefficient (violet) (b, d) are derived from the SE fit (solid lines) and the T/R fit (dash-dotted lines).

For determining the linear optical constants, each heterostructure was considered as an effective medium, as opposed to a layer-by-layer approach. The Drude-Lorentz oscillator model was applied to fit the SE data and the optical constants were subsequently determined. The number of oscillators used in the models for each composite varied from 2 to 7 depending on the complexity of the system. Heterostructures with 32 and 64 Ir ALD cycles (i.e., ≈ 2 nm and 4 nm, respectively) were found to be the most challenging to model, by virtue of these systems being in the transition regime, whereas modeling the heterostructures with an unambiguous dielectric or metallic character was more straightforward. Once an acceptable fit of the SE data was attained, the T and R spectra were simulated from the oscillator model and compared to the measured T/ R values. The fits of the SE data were only considered satisfactory if the simulated T, R values reasonably matched the measured spectra. An example of the fitting procedure for two composites is given in Figure 7. The measured T/R spectra with the corresponding OL are shown with solid curves and the T/R/OL spectra derived from the fits of the SE measurements are shown in dashed curves. The wavelength range of the SE measurements was 200-980 nm; the derived T/R/OL spectra consequently span a narrower spectral range than the T/R measurements, which were performed from 200 - 2000 nm wavelength range. Furthermore, the measured T/R data were also fitted with the Drude-Lorentz oscillator model, as shown with dots in Figure 7 (a, c). The optical constants were then derived from the T, R fits, generating a second set of n and k values in addition to the values derived from the SE fits. A comparison of the two data sets for *n* and *k* is shown in Figure 7 (b, d), with the values from the SE fit in solid lines and the ones from the T/R fit in dash-dotted lines. Very good agreement between the two sets increases our confidence in the reliability of the models for the optical constants presented in the following discussion. Discrepancies might also be attributed to slight variations due to different substrates and the nonuniformity of film growth.

The refractive index profiles of heterostructures with thinner Ir layers (≈ 1 nm, up to 16 cycles) resemble that of a single layer $Al_2O_3$ film; in both cases with $Al_2O_3$ spacer layers of 35 or 15 ALD cycles, as illustrated in Figure 8 (a, b) and (c, d), respectively. According to the dielectric behavior of these heterostructures, there is no significant extinction up to 16 Ir cycles, except for a slightly higher edge in the UV spectral range compared to the extinction of a single-layer $Al_2O_3$ film. With an increasing cycle number of Ir, a gradual increase in the refractive index *n* and the extinction coefficient *k* is observed since the metallic character starts predominating as the number of Ir cycles approaches 32. The extinction profile attains distinct enhancement along with increasing metallic concentration. The set of heterostructures with thinner spacer layers (see Figure 8. (c, d)) gave rise to stronger extinction than its thicker counterparts (see Figure 8. (a, b)). The dispersion of *k* exhibits a significant metallic character in the stacks with 64 Ir ALD cycles; this metallic character is more pronounced for the thinner spacer layer systems. However, the critical Ir contribution for the effective insulator-to-metal transition is within 32 to 64 cycles of Ir (≈ 2 - 4 nm). Finally, at 128 Ir cycles (≈ 7.6 nm), *n* and *k* possess similar trends as that of a pristine Ir layer.

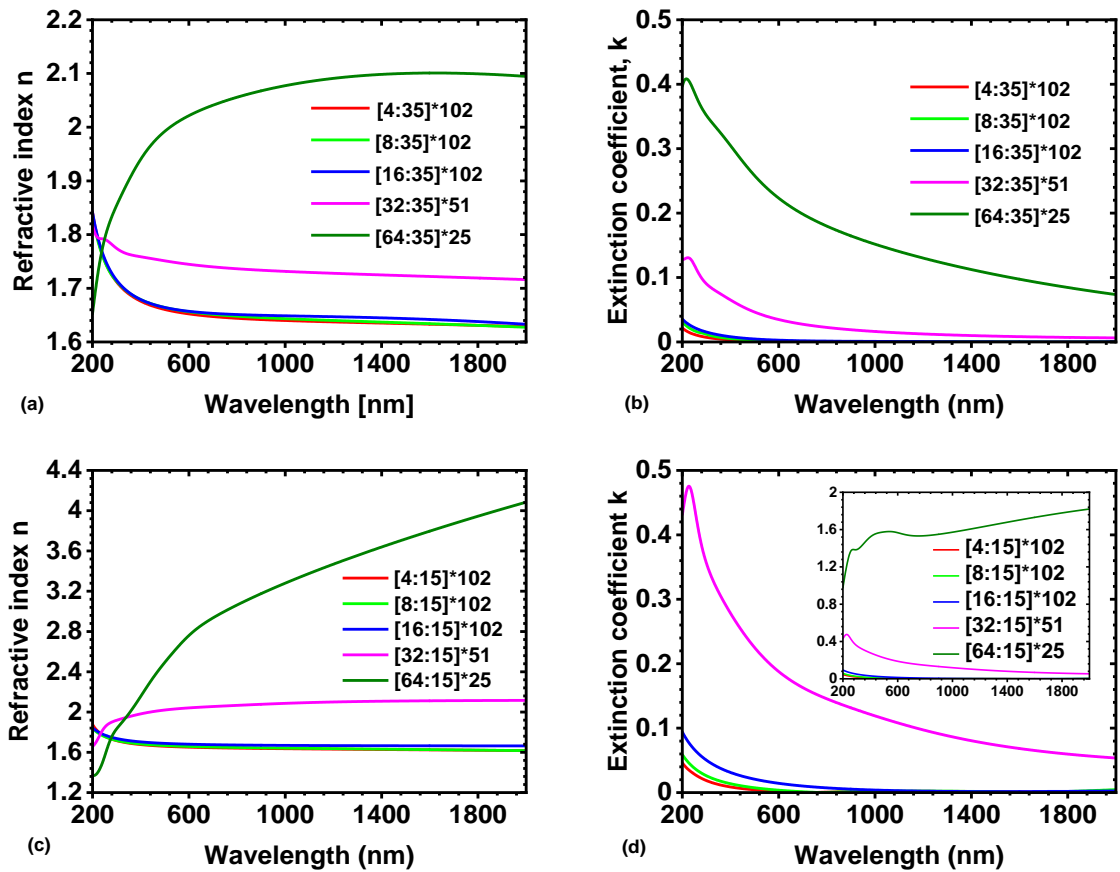

**Figure 8.** Refractive index *n* and extinction coefficients *k* of Ir/Al$_2$O$_3$ heterostructures with (a, b) 35 cycles and (c, d) 15 cycles of Al$_2$O$_3$ spacer layers.

The transition regime, meaning, the regime where the heterostructures undergo an effective insulator to metal transition as the composition ratio varies, becomes evident when considering the effective dielectric function ($\varepsilon$) of the heterostructures. The real and imaginary parts of $\varepsilon$ in the UV/VIS spectral range are shown in Figure 9 (a-d). The real part of the dielectric function Re($\varepsilon$) varies from being completely positive to completely negative across this entire spectral range. Noteworthy is the composition ratio at which the system is exactly in the transition regime, i.e., the [64:15]*25 heterostructure. As shown in Figure 9 (a), Re($\varepsilon$) for this structure approaches zero at around 230-240 nm wavelength. In materials where Re($\varepsilon$) crosses zero, a strong non-linear optical response can be observed at the epsilon-near-zero (ENZ) frequencies [38,39]. Therefore, the [64:15]*25 heterostructure containing Ir and Al$_2$O$_3$ interfaces can be a potential candidate for applications in the field of non-linear optics. However, the imaginary part of the dielectric function Im($\varepsilon$) is relatively high, around 4.2 to 4.6. The [64:15]*25 heterostructure also exhibits high optical losses of roughly around 90% in its ENZ wavelength range, as seen in Figure 6(c). High optical losses remain one of the largest obstacles faced in ENZ material development, which has led to criticism of the practicality of ENZ materials [40]. Several means have been attempted to compensate for the high optical losses observed in many ENZ materials, for example, by combining the ENZ material with dielectrics in a multilayer structure [41-43] or by doping the ENZ material with a gain medium [44]. Here, reducing film thickness from approximately 100 nm to several tens of nm would also result in lower absorption losses based on Lambert-Beer's law.

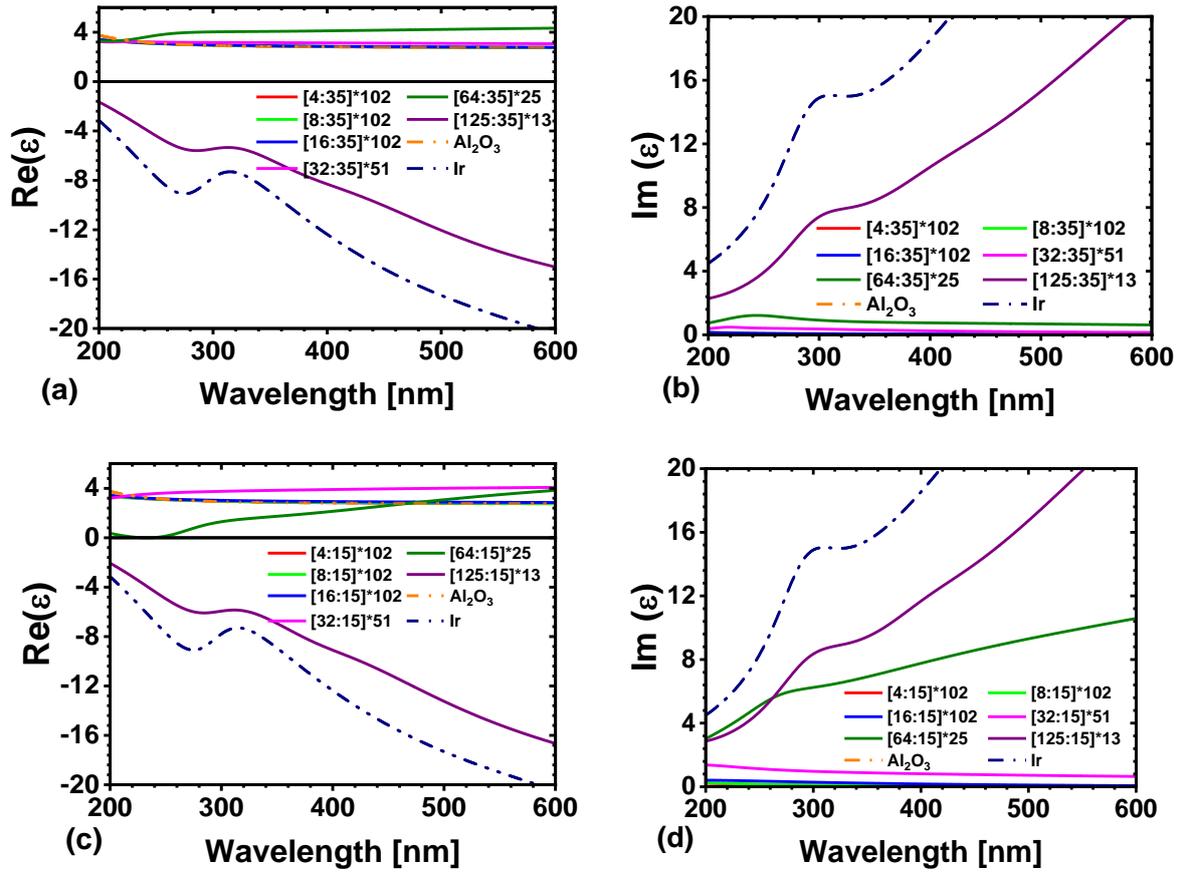

**Figure 9.** The real part Re($\epsilon$) and imaginary part Im($\epsilon$) of the dielectric function of Ir/Al$_2$O$_3$ heterostructures with (a, b) 35 cycles and (c, d) 15 cycles of Al$_2$O$_3$ spacer layers.

Additionally, effective medium approximation (EMA) approaches have been implemented to simulate the optical constants of these heterostructures. We have applied the Maxwell–Garnett (MG) and Bruggeman (BG) formalisms to numerically obtain a rough estimation of the effective *n* and *k* spectra of the heterostructures. The MG model is based on including a material in a host matrix. It breaks down at a higher volume fraction of the inclusion material as the particle interaction is not considered, and the model loses its validity. In contrast, the BG framework considers the system to be a uniform mixture of two constituent materials. Figure 10 shows the effective *n* and *k* employing the MG and BG models for various amounts of volume fractions of Ir in the Ir/Al$_2$O$_3$ heterostructures. A case study is shown here with compositions having 35 cycles of the Al$_2$O$_3$ spacer layer. Considering the MG model (see Figure 10 (a, b)), an increasing trend can be seen in the optical constants as metallic influence increases; however, the model deviates as the volume fraction approaches 50%. As expected, the BG model (see Figure 10 (c, d)) demonstrates a better approximation at higher Ir volume fractions than the MG model. With increasing Ir content in the BG model, the trend of the optical constants becomes increasingly more comparable to a pristine Ir layer. Nevertheless, the *n* profile could not follow the experimental observations for the heterostructure with the thickest Ir layers, (i.e., 128 cycles ≈ 7.6 nm), which can be attributed to the formation of closed Ir layers in Ir/Al$_2$O$_3$ nanolaminate systems leading to the unsuitability of the mixture model assumptions.

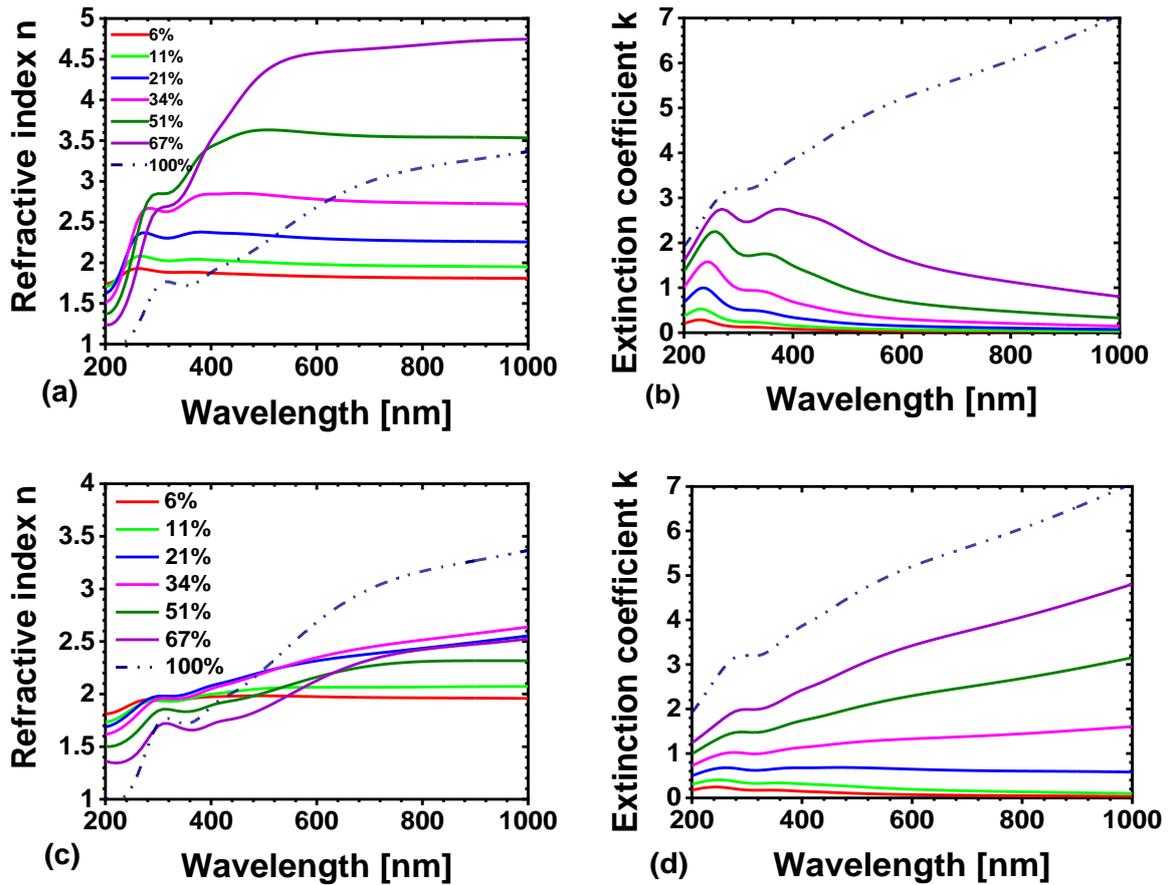

**Figure 10.** Refractive index *n* and extinction coefficients *k* of Ir/Al$_2$O$_3$ heterostructures, simulated using (a, b) Maxwell-Garnett (MG) and (c, d) Bruggeman (BG) models. The simulations are performed for different volume fractions of Ir corresponding to the Ir/Al$_2$O$_3$ heterostructures mentioned in Table 1, all having 35 cycles of the Al$_2$O$_3$ spacer layer.

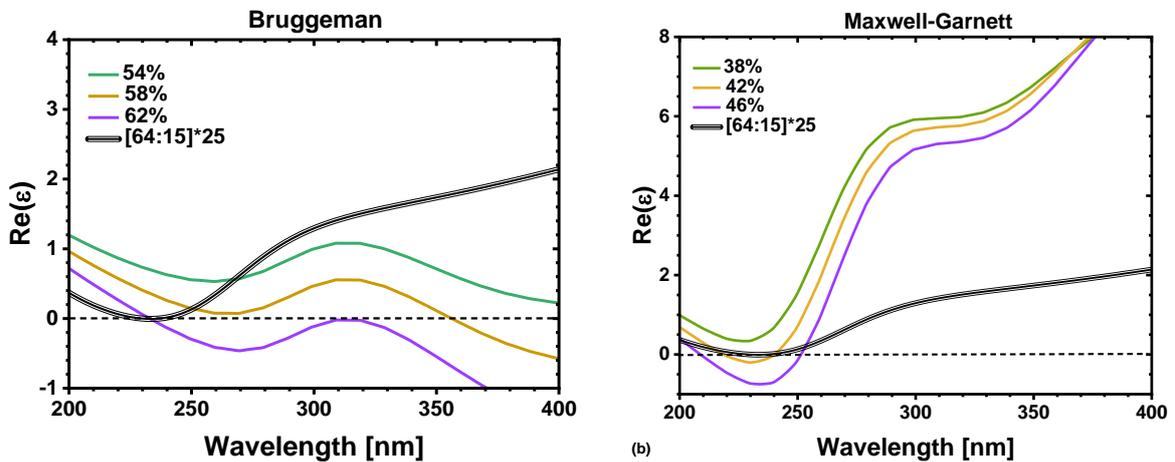

**Figure 11.** Real part Re($\epsilon$) of the dielectric function calculated according to two effective medium approximation (EMA) approaches; (a) Bruggeman and (b) Maxwell-Garnett. The colored lines indicate various volume fractions of Ir in each composite model which show resemblance to dispersion behaviour of [64:15]*25 heterostructure near the ENZ wavelength region. The black (thick) curve represents the dielectric function of the [64:15]*25 heterostructure, derived from the fits of SE measurement.

The optical constants from the EMA models and fabricated heterostructures were also compared, with particular attention given to the [64:15]*25 heterostructure due to its ENZ behavior. Figure 11 depicts the real dielectric function of [64:15]*25 as derived from the SE measurement fits, alongside with the EMA models for selected amounts of Ir volume fractions. The interesting wavelength region was considered between 200 nm and 400 nm targeting the ENZ regime of the [64:15]*25 heterostructure. The volume fractions shown in each graph are the ones that bear some resemblance to Re($\varepsilon$) of the fabricated heterostructure, although neither the BG model nor the MG model was able to accurately predict the behavior of the dielectric function of the [64:15]*25 heterostructure. Material models with higher volume fractions of Ir are comparable to the Re($\varepsilon$) of [64:15]*25 when focusing on 200-400 nm wavelength, taking into account that the Ir content of this fabricated heterostructure is roughly 50%. However, considering the entire spectral range, the material models that estimate the closest match to the [64:15]*25 composite have a significantly lower Ir contribution than expected.

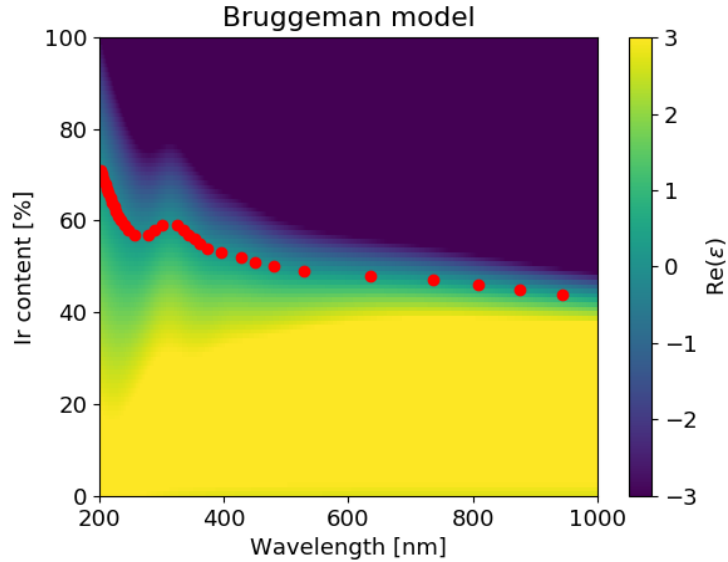

**Figure 12.** Real part Re($\epsilon$) of the dielectric function $\epsilon$ simulated according to the Bruggeman effective medium approximation approach. The volume fraction of Ir was varied from 0% to 100%.

Figure 12 demonstrates the simulated ENZ wavelengths of Ir/Al$_2$O$_3$ heterostructures with varying volume fractions of Ir. The red dots represent the ENZ wavelengths for the corresponding compositions. Furthermore, a color gradient indicating the value of Re($\varepsilon$) was included to show the range where Re($\varepsilon$) is approaching zero. In the wavelength range of 200 -1000 nm, ENZ occurs for heterostructures with 40 to 80% Ir content. This emphasizes on how small changes in the composition ratio result in major shifts in the ENZ wavelength. A precise tunability of the composition ratio is thus crucial to achieve the ENZ behavior at a targeted wavelength. It should be noted that this material model does not account for the possibility of IrO$_x$ forming at Ir/Al$_2$O$_3$ interfaces. Even though these simulations cannot precisely estimate the behavior of such heterostructures, they provide a reliable trend of Re($\varepsilon$) with increasing volume fraction of the metal. Further improvement of material modeling will be the scope of future research to facilitate improved material design and fabrication. Considering these challenges, an advantage of an Ir/Al$_2$O$_3$ heterostructures as ENZ material is the high tunability of the effective dielectric function. As the metal-insulator transition is relatively rapid, the dielectric function varies significantly with minor changes in the number of ALD cycles. The Ir/Al$_2$O$_3$ heterostructures could be optimized for implementations as ENZ metamaterials in nonlinear optical applications by precisely tuning the composition enabled by ALD.

**CONCLUSIONS**

We developed a potential route to fabricate atomic-scale metal-dielectric heterostructures of Ir/Al$_2$O$_3$ by the ALD method. A combination of spectroscopic and microscopic methods was implemented to understand the interfacial properties thoroughly. XRR and HRTEM studies reveal the existence of layered heterostructures in contrast to island growth expected for metals on dielectrics. Furthermore, our XPS analysis support the formation of an Ir-O-Al environment for Angstrom scale Ir contributions. We observed a precisely controlled growth of an atomically thin Ir metallic layer ($\approx$ 68% metallicity) starting from about 1 nm. A holistic and reliable set of characterizations was developed to determine the thickness and dispersion of such Ir/Al$_2$O$_3$ heterostructures. The linear optical properties demonstrate the possibility of tuning the optical constants, i.e., eventually the optical band structure, by precise tuning of compositions. Furthermore, manipulating the dielectric function enables to obtain ENZ metamaterials based on metal-dielectric multilayer films. The precise tuning of composition provided by the ALD technology can tailor the ENZ condition. Overall, these observations pave the path to explore new optical materials with an engineered structural and optical properties for various applications.

## AUTHOR INFORMATION

**Corresponding Author**

*E-mail: adriana.szeghalmi@iof.fraunhofer.de

**Notes**
**The authors declare no conflict of interest.**

**Funding Sources**

This research was supported by the German Research Foundation (DFG, Deutsche Forschungsgemeinschaft) Collaborative Research Center (CRC/SFB) 1375 "NOA - Nonlinear Optics down to Atomic scales" – project B3, number 398816777 and the Fraunhofer Society Attract Project (grant number 066-601020), Fraunhofer IOF Center of Excellence in Photonics.

## ACKNOWLEDGMENT

The authors gratefully acknowledge David Kästner for the technical support and Ingo Uschmann for the GIXRD measurements.